\documentclass[letter]{aa}     % for the letters
\usepackage{graphicx}
\usepackage{adjustbox}
\usepackage{txfonts}
\usepackage{mhchem}
\usepackage{lipsum}
\usepackage{subcaption}         % necessary for continued figures, example in section 3
\usepackage{lscape}             % to rotate a single page table, example in appendix.
\usepackage{placeins}           % useful with \FloatBarrier, to keep 
\newcommand{\rev}[1]{#1}                                
\usepackage{hyperref}
\begin{document}

\title{First detection of s-cis-acroleine (\ce{sc-CH2CHCHO}) in the interstellar medium}
%\subtitle{An abundant isomer }

\author{G. Molpeceres\inst{1}
\and M. Ag\'undez\inst{1}
\and C. Cabezas\inst{1}
\and N. Marcelino\inst{2,3}
\and G. Esplugues\inst{2,3}
\and B. Tercero\inst{2,3}
\and J. Cernicharo\inst{1}}

\institute{
Departamento de Astrofísica Molecular,
Instituto de Física Fundamental (IFF-CSIC),
C/ Serrano 121, 123, 113b, 28006 Madrid, Spain
\email{german.molpeceres@iff.csic.es, jose.cernicharo@csic.es}
\and
Observatorio Astronómico Nacional, IGN,
Calle Alfonso XII 3, E-28014 Madrid, Spain 
\and
Observatorio de Yebes (IGN). Cerro de la Palera s/n, 19141 Yebes, Guadalajara, Spain
}

\date{Received; accepted}

% Structured abstract: context, aims, methods, results, conclusions.
\abstract{
We report the first interstellar detection of the higher-energy $s$-cis conformer of acrolein (\ce{CH2CHCHO}) toward TMC-1 using the \textsc{Quijote} line survey. Local thermodynamic equilibrium modeling, assuming a rotational temperature of 7.5~K, yields a column density of $(1.0\pm0.2)\times10^{11}$~cm$^{-2}$ when fitting the observed lines. The $s$-cis/$s$-trans abundance ratio is $0.45\pm0.11$, despite an energy separation of approximately 1025~K, indicating a strongly nonequilibrium conformational distribution. An exploratory gas--grain chemical model identifies \ce{C3H5 + O} as the dominant acrolein formation route, pointing to this reaction as a first entry to consider when explaining the isomeric ratios. The observed ratio provides a new constraint on conformer-specific formation, destruction, and interconversion processes in cold interstellar clouds.
}


\maketitle

\section{Introduction}

The detection of interstellar isomers, molecules with the same chemical formula but different structural arrangements, provides valuable insights into the chemistry of astronomical environments. Their abundance ratios can serve as probes of the physical and chemical conditions of the regions where they are observed \citep[e.g.,][]{goicoechea_ionization_2009,cuadrado_trans-cis_2016,hacar_hcn--hnc_2020,loison_interstellar_2017}. Particularly sensitive to these conditions are stereoisomers, which share the same atomic connectivity but differ in their three-dimensional arrangement. Their relatively small energy differences and largely similar chemistry make their abundance ratios highly sensitive to even modest changes in physical or chemical conditions \citep{cuadrado_trans-cis_2016,Shingledecker2020,shingledecker_grain-surface_2022,molpeceres_hydrogen_2022,garcia_de_la_concepcion_sequential_2023,sanz-novo_conformational_2025,molpeceres_formic_2025}. A recent review of interstellar isomerism is provided by \citet{rivilla_interstellar_2026}.

\begin{figure}
    \centering
    \begin{minipage}{0.45\linewidth}
        \centering
        \includegraphics[width=\linewidth]{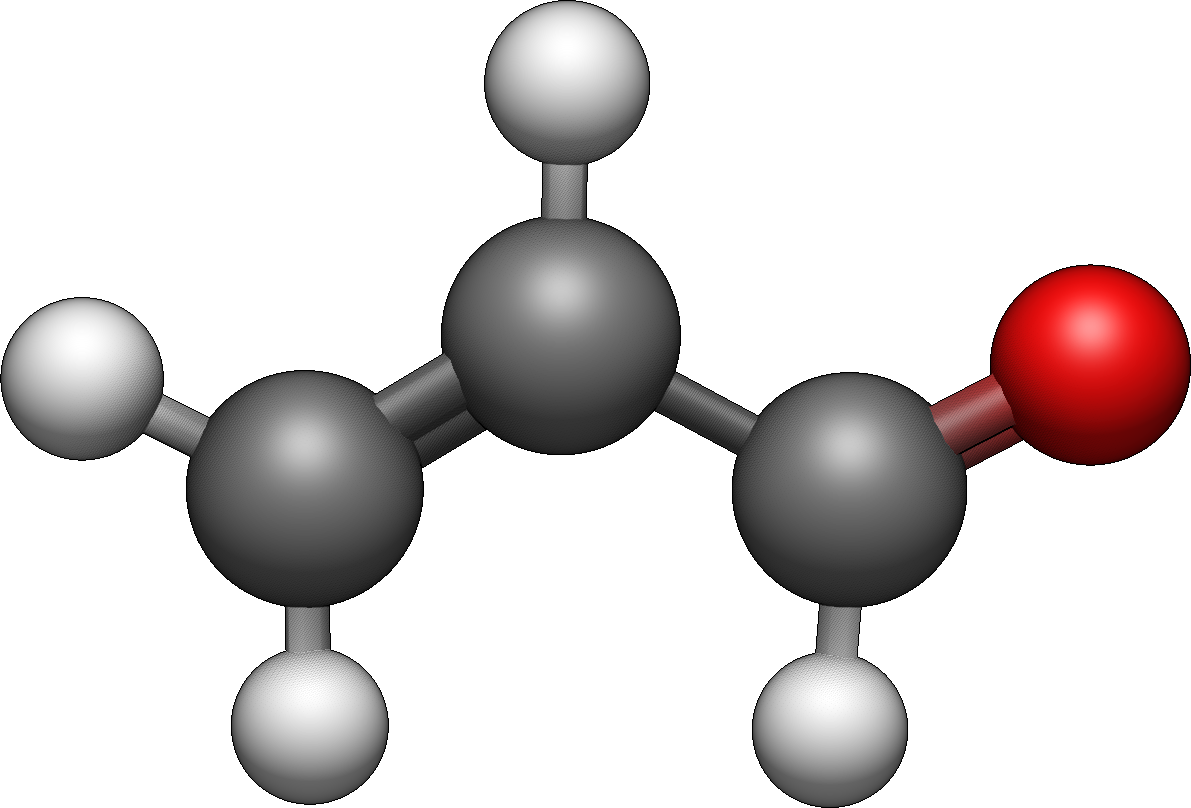}
        \Large
        \textit{s-trans}
    \end{minipage}
    \hspace{0.4in}
    \begin{minipage}{0.40\linewidth}
        \centering
        \includegraphics[width=\linewidth]{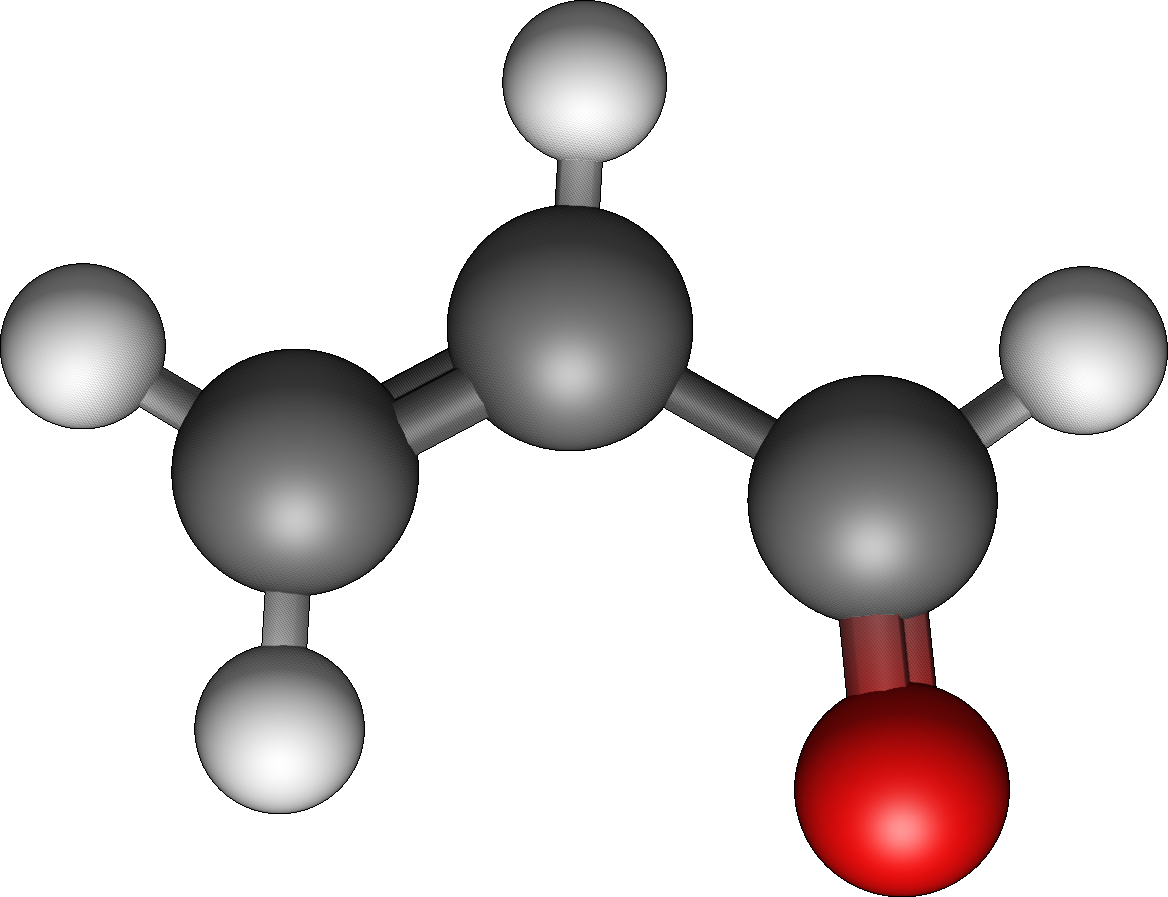}
        \Large
        \textit{s-cis}
    \end{minipage}
    \caption{Molecular structures of the \textit{s-trans} and \textit{s-cis} conformers of acroleine (\ce{CH2CHCHO}). The separation between both conformers is 712 cm$^{-1}$ (approximately 1025 K, \citealt{puzzarini_molecular_2014}). }
    \label{fig:species}
\end{figure}

The contemporary observations conducted by the \textsc{Quijote} and \textsc{Gotham} surveys\footnote{QUIJOTE: Q-band Ultrasensitive Inspection Journey to the Obscure TMC-1 Environment \citep{cernicharo_quijote_2022}; GOTHAM: GBT Observations of TMC-1: Hunting Aromatic Molecules \citep{mcguire_early_2020}.} have led to the detection of several stereoisomers in TMC-1  \citep[e.g.,][]{Lee2021,cernicharo_discovery_2022,cabezas_discovery_2025,molpeceres_formic_2025}. In particular, using \textsc{Quijote} observations, \citet{agundez_o-bearing_2021} reported the detection of the s-trans conformer\footnote{Following \citet{rivilla_interstellar_2026}, we adopt the IUPAC term ``conformer'' for stereoisomers that interconvert through rotation around a single bond, and the prefixes ``st'' and ``sc'' to distinguish the s-trans and s-cis forms, respectively (see \url{https://goldbook.iupac.org/terms/view/S05508}).} of acrolein (\ce{C2H3CHO}), \rev{also known as propenal}. \rev{Thanks to the improved sensitivity achieved during the final observations of the survey, we report here the detection of acrolein’s higher-energy \(s\)-cis conformer} The two conformers are shown in Fig \ref{fig:species}. They are separated by an energy difference of 712~cm$^{-1}$, a small gap by terrestrial chemical standards, but enormous for interstellar chemistry. The detection of both conformers thus provides a unique opportunity to investigate the chemical processes governing conformational populations under the extreme conditions of the interstellar medium.

\section{Observations}

The detection of s-cis-\ce{CH2CHCHO} is framed in the context of the \textsc{Quijote} line survey of the TMC-1 cloud \citep{Cernicharo2021, cernicharo_discovery_2026}. This survey has been carried out with the Yebes 40m telescope and covers the $\mathcal{Q}$ frequency band (31--50 GHz) and is centered on the cyanopolyyne peak of the source ($\alpha_{\rm J2000} = 4^{h}41^{m}41.9^{s}$, $\delta_{\rm J2000} = +25^{\circ}41'27''$), one of the most chemically rich regions of the interstellar medium (ISM). The \textsc{Quijote} observations were conducted in frequency-switching mode, using frequency throws of 8 or 10 MHz. Observations were carried out using a 7 mm dual-linear-polarization receiver (HEMT) coupled to eight fast Fourier transform spectrometers, enabling simultaneous coverage of the entire $\mathcal{ Q}$ band with a spectral resolution of 38.15 kHz in both polarizations \citep{tercero_yebes_2021}. Receiver temperatures
are 16 K at 32 GHz, and 30 K at 50 GHz. The beam efficiency in the full analysed band can be obtained as $B_{\rm eff}=0.797 \exp\left[\left(-\nu/71.1\right)^2\right]$ with \rev{$\nu$} in GHz. The forward telescope efficiency is 0.97.
The telescope beam size at half power intensity is 54.4'' at
32.4 GHz and 36.4'' at 48.4 GHz. At the time of writing \textsc{Quijote} accumulates 1509.2 hours of integration time, roughly equally divided for data with 8 and 10 MHz frequency throws. For the combined data, the sensitivity of \textsc{Quijote} varies roughly between 0.06--0.18 mK depending on the specific frequency \citep[for more details, check Figure 2 of][]{agundez_tmc-1_2026}. The analysis of the spectral data is carried out with the \textsc{Gildas} suite, using the \textsc{Class} program\footnote{\url{http://www.iram.fr/IRAMFR/GILDAS}} in conjunction with the spectral modeling capabilities of the \textsc{Madex} code \citep{cernicharo_laboratory_2012}\footnote{https://nanocosmos.iff.csic.es/madex/} that incorporates the spectroscopic information of the molecule \citep{blom_microwave_1982,blom_molecular_1984, daly_comprehensive_2015, mccarthy_exhaustive_2020}. 

\section{Results}

\begin{figure*}
    \centering
    \includegraphics[width=0.9\textwidth]{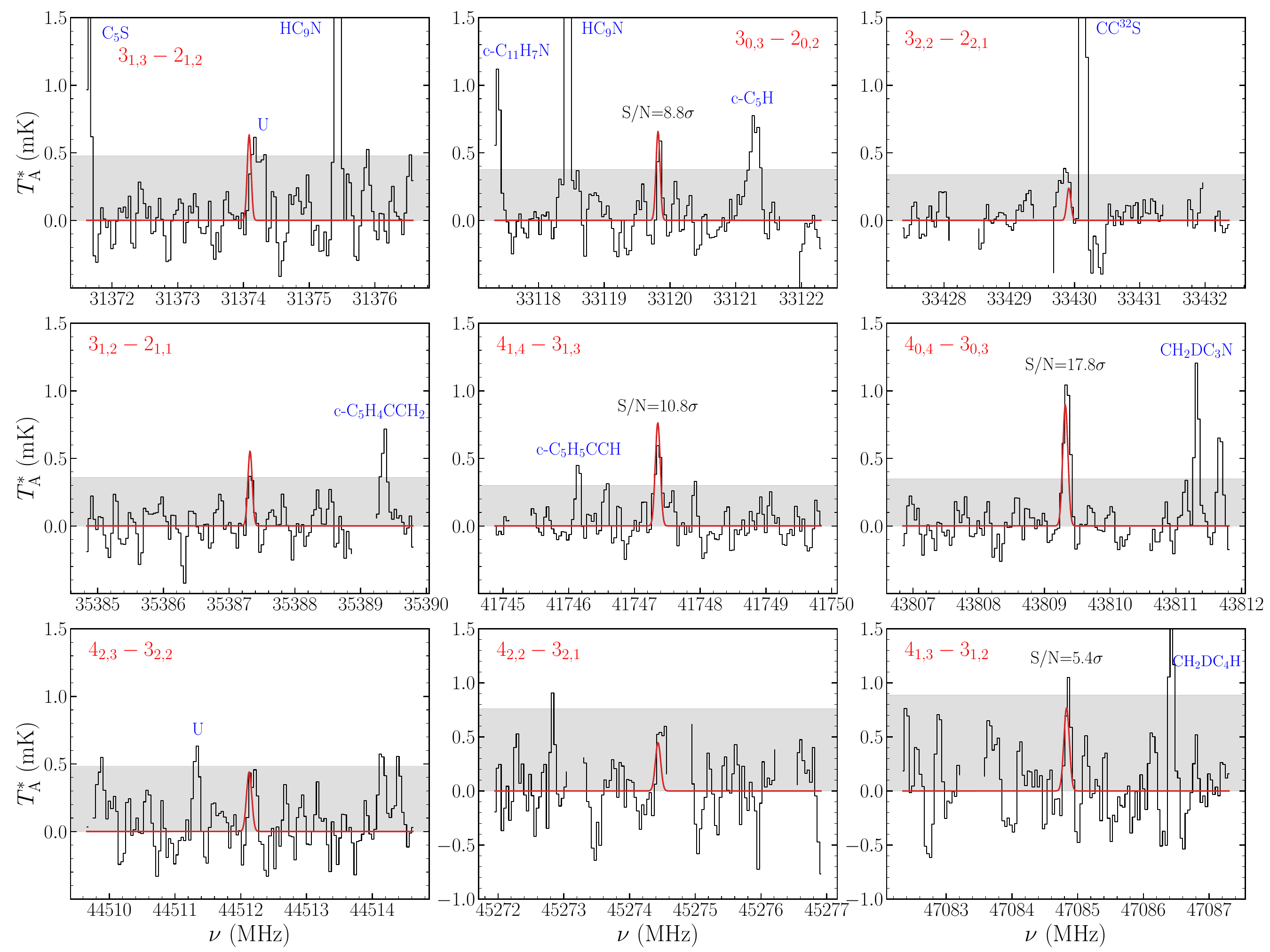}
    \caption{Most intense s-cis-\ce{CH2CHCHO} predicted transitions assuming local thermodynamic equilibrium at 7.5 K in continuous red line along with the corresponding window of the \textsc{Quijote} line survey. In blue, assigned lines besides the ones of s-cis-\ce{CH2CHCHO}. \rev{Blanked channels correspond to negative artifacts of the frequency switching technique. The shaded area represents the 3$\sigma$ level in the shown frequency range of each panel, where $\sigma$ is the RMS measured at the nominal spectral resolution of 38.15 kHz ($\delta\nu$). For molecules detected above 3$\sigma$ we also report the S/N level calculated as \(S/N=\int T_A^*\,dv/\left(\sigma\sqrt{\Delta v\,\delta\nu\,c/\nu}\right)\) \citep{agundez_o-bearing_2021} where all magnitudes are given in Table \ref{tab:lines} and this caption, except $c$, the velocity of light.}}
    \label{fig:spectra}
\end{figure*}

\begin{table}
\caption{Observed line parameters of s-cis-\ce{CH2CHCHO}. Uncertainties are estimated from a Gaussian fit of the transition using \textsc{Class}, unless stated otherwise.}
\label{tab:lines}
\centering

\begin{adjustbox}{max width=\columnwidth}
\begin{tabular}{lcccccc}
\hline\hline
Transition &
$\nu$ &
$T_{\rm A}^{*}$ &
E$_{\rm upp}$ &
$\Delta v$ &
v$_{\rm LSR}$ &
$\int T_{\rm A}^{*}dv$ \\
&
(MHz) &
(mK) &
(K) &
(km\,s$^{-1}$) &
(km\,s$^{-1}$) &
(mK\,km\,s$^{-1}$) \\
\hline

3$_{1,3}$--2$_{1,2}$$^{a}$ &
31374.086 & 0.55 & 3.8 & N/A & N/A & N/A \\

3$_{0,3}$--2$_{0,2}$ &
33119.818 & \rev{0.61} & 3.2 & 0.74$\pm$0.18 & 5.59$\pm$0.07 & 0.56$\pm$0.11 \\

3$_{2,2}$--2$_{2,1}$$^{a}$ &
33429.867 & \rev{0.39} & 6.5& N/A & N/A & N/A \\

3$_{1,2}$--2$_{1,1}$ &
35387.318 & 0.42 & 4.2 & 0.77$\pm$0.21 & 5.74$\pm$0.10 & 0.34$\pm$0.09 \\

4$_{1,4}$--3$_{1,3}$$^{b}$ &
41747.355 & 0.57 & 5.8 & \rev{0.6} & \rev{5.75$\pm$1.0} & \rev{0.44$\pm$0.87}\\

4$_{0,4}$--3$_{0,3}$ &
43809.318 & 1.10 & 5.3 & 0.79$\pm$0.13 & 5.68$\pm$0.06 & 0.94$\pm$0.15 \\

4$_{2,3}$--3$_{2,2}$$^{c}$ &
44512.127 & 0.39 & 8.7 & 0.51$\pm$0.24 & 5.60$\pm$0.11 & 0.21$\pm$0.09 \\

4$_{2,2}$--3$_{2,1}$$^{d}$ &
45274.434 & 0.60 & 8.7 & N/A & N/A & N/A \\

4$_{1,3}$--3$_{1,2}$ &
47084.832 & 1.03 & 6.5 & 0.54$\pm$0.20 & 5.69$\pm$0.10 & 0.58$\pm$0.21 \\
\hline
\end{tabular}
\end{adjustbox}

\tablefoot{
$\nu$ is the rest frequency;
$T_{\rm A}^{*}$ is the peak antenna temperature;
$\Delta v$ is width of the line;
$v_{\rm LSR}$ is the radial velocity; and
$\int T_{\rm A}^{*}dv$ is the velocity-integrated antenna
temperature. $^{a}$- A reliable fit to the line could not be obtained due to \rev{significant blending.}  $^{b}$-\rev{Fit with fixed $\Delta\nu$ and large uncertainty in the area}. \rev{$^{c}$- Partially blended}. $^{d}$- No fit is attempted due to the presence of an artifact of the frequency switching technique close to the line (See Figure \ref{fig:spectra})
}
\end{table}

The a-component of the electric dipole moment of sc-\ce{CH2CHCHO} ($\mu_a$=2.01 Debye) is slightly lower than that of the low energy conformer st-\ce{CH2CHCHO} ($\mu_a$=3.05 Debye) \citep{blom_microwave_1982, blom_molecular_1984}. Therefore, the lines of the s-cis conformer are weaker than those of the s-trans conformer, by a factor of 2.3 due to the difference in dipole moments plus an additional factor due to the lower abundance (see below). Serving as an example in \citet{agundez_o-bearing_2021} the detected lines ranged in the 2--5 mK level whereas in this work the most intense line peaks at around 1 mK. Nevertheless, owing to the high sensitivity of the \textsc{Quijote} survey, we were able to firmly establish the detection of sc-\ce{CH2CHCHO} through a line-by-line analysis. The strongest predicted transitions, which correspond to the most prominent observed features, are presented in Figure~\ref{fig:spectra} and compiled in Table~\ref{tab:lines}. 

All detected transitions belong to the $J_{\rm upp}$=4,3 and $J_{\rm low}$=3,2 families of lines, with some of them being partially blended. All unblended transitions exhibit narrow line profiles, with widths ($\Delta \nu$ in Table \ref{tab:lines}) consistent with the characteristic full width at half maximum (FWHM) of $\sim$0.6 km s$^{-1}$ observed toward TMC-1 \citep{cernicharo_tmc-1_2021, fuentetaja_analysis_2025, cabezas_discovery_2025} and all lines appear consistently aligned with the systemic velocity of the source $v_{\rm LSR}$=5.83 km s$^{-1}$ \citep{2020A&A...642L...8C}. Most notably, we identify \rev{four} transitions detected above the 3$\sigma$ noise level that appear to be entirely unblended, namely the 3$_{0,3}$--2$_{0,2}$, 4$_{0,4}$--3$_{0,3}$, and 4$_{1,3}$--3$_{1,2}$ ones, \rev{or only slightly blended, as is the case of the 4$_{1,4}$--3$_{1,3}$ one}. We attempted a rotational diagram analysis using them, however, significant uncertainties in the determination of $\int T_{\rm A}^{*}dv$ prevented us from reliably constraining the rotational temperature and the column density ($N$). To determine $N$, we therefore followed the approach adopted in some of our previous detections \citep{cernicharo_pure_2021, cernicharo_discovery_2021, esplugues_first_2025}, fitting synthetic spectra generated with the \textsc{Madex} code directly to the observed line profiles, rather than constructing a rotational diagram. Briefly, we used the \textsc{Madex} code \rev{to derive in absence of collisional rates, the} local thermodynamic equilibrium (LTE) synthetic spectrum of sc-\ce{CH2CHCHO} (overlayed red line in Figure \ref{fig:spectra}). The rotational temperature was assumed to be 7.5 K as the one used in \citet{agundez_o-bearing_2021} for st-\ce{CH2CHCHO}. The diameter of the source is assumed as 80'' \citep{fosse_molecular_2001, cernicharo_spatial_2023} and the line width is set to 0.6 km s$^{-1}$ as explained above. We assume an uncertainty in the column densities determined through this protocol of $\pm$20\% \citep{marcelino_first_2023,esplugues_first_2025}. From this procedure we determine a best fit column density corresponding to $N$=1.0$\pm$0.2$\times$10$^{11}$ cm$^{-2}$ for sc-\ce{CH2CHCHO} which translates in a fractional abundance $n_{\ce{sc-CH2CHCHO}}$/$n_{\ce{H2}}$=1.0$\pm$0.2$\times$10$^{-11}$ for a $N_{\ce{H2}}$=1.0$\times$10$^{22}$ \rev{cm$^{-2}$} \citep{1987A&A...176..299C}.

\section{Discussion}

The detection of sc-\ce{CH2CHCHO} is best understood in the broader context of interstellar stereoisomerism. The column density previously derived for the lower-energy conformer st-\ce{CH2CHCHO} is $(2.2\pm0.3)\times10^{11}$ cm$^{-2}$ \citep{agundez_o-bearing_2021}, yielding an st/sc abundance ratio of $2.2\pm0.5$. This yields an observed Stereoisomeric Ratio (OSR, defined as $N_{\rm high}/N_{\rm low}$; \citealt{rivilla_interstellar_2026}) of $0.45\pm0.11$. This value contrasts sharply with the calculated Boltzmann factor at 9 K ($\Delta$E=1025 K) of $\sim10^{-50}$, indicating that the relative abundances are kinetically controlled, as expected at the extremely low temperatures of TMC-1. Below, we discuss this result in the context of interstellar stereoisomerism and explore possible explanations for the unexpectedly large sc/st abundance ratio.

\begin{table}[t]
\centering
\caption{Observed stereoisomeric ratios (OSRs) in TMC-1.}
\label{tab:stereoisomers}

\resizebox{\columnwidth}{!}{%
\begin{tabular}{lccc}
\hline\hline
Molecule & OSR & $\Delta E$ & Reference \\
\hline
c/t-HCOOH & 0.06 & 2050 & \citep{molpeceres_formic_2025} \\
g/c-\ce{CH2CHCH2CN} & 1.14 & 1143 & \citep{cernicharo_discovery_2022} \\
c/t-\ce{CH3CHCHCN} & 2.60 & 96 & \citep{cernicharo_discovery_2022} \\
sc/st-\ce{CH2CHCHO} & 0.45 & 1025  & This work \\
sc/st-\ce{CH2CHCHS}\textsuperscript{1} & $<$2.93 & 1207 & \citep{molpeceres_laboratory_2026} \\
t-Z/t-E-\ce{HCCCHCHCN}\textsuperscript{1} & $<$1.00 & $\sim$240 & \citep{Lee2021} \\
\hline
\end{tabular}%
}
\tablefoot{\rev{$\Delta E$ is the energy gap between isomers obtained from the original references in Kelvin}. The OSR is defined as the column-density ratio of the higher-energy stereoisomer to the lower-energy stereoisomer, following \citep{rivilla_interstellar_2026}. \textsuperscript{1} The sc-\ce{CH2CHCHS} and t-Z-\ce{HCCCHCHCN} isomers are not detected; upper limits for the OSRs are calculated from the reported upper limits. In \citet{molpeceres_laboratory_2026}, we used the syn/anti nomenclature. Here, we use the IUPAC-recommended nomenclature for consistency (see footnote 2 in the Introduction). }
\end{table}

\begin{figure}
    \centering
    \includegraphics[width=0.9\linewidth]{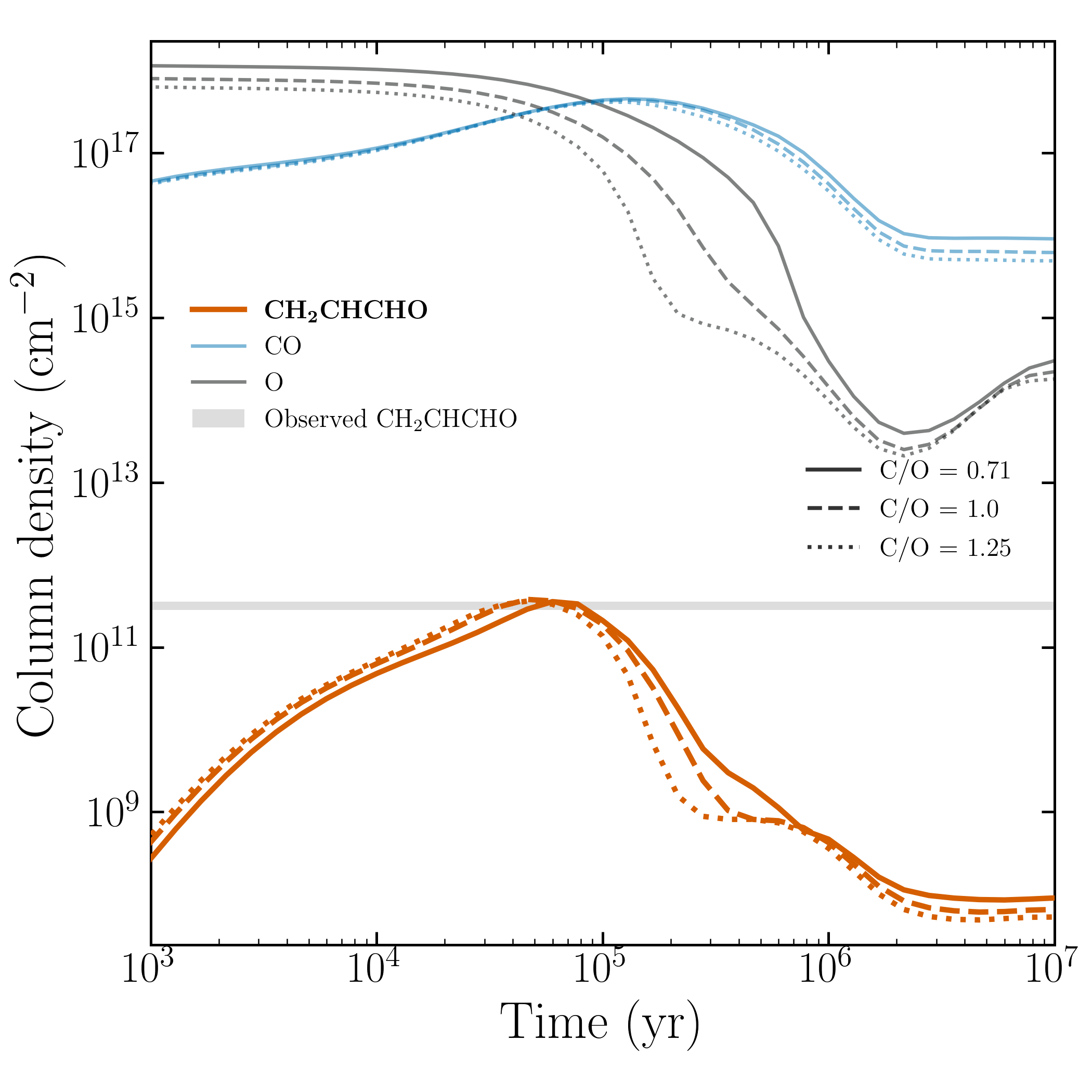}
    \caption{Chemical model for the formation of \ce{CH2CHCHO}. We note that the reactions added in \citet{molpeceres_laboratory_2026} Appendix B,for \ce{CH2CHCHS} are not included in this model.}
    \label{fig:model}
\end{figure}

The molecule investigated here, sc-\ce{CH2CHCHO}, is the fourth stereoisomer detected in TMC-1. Table~\ref{tab:stereoisomers} summarizes the OSRs of previously detected stereoisomeric pairs, including upper limits where only one conformer has been detected. Except for HCOOH, all three pairs of detected conformers exhibit OSRs near or above unity. Recent theoretical calculations suggest that gas-phase formation pathways largely explain the observed OSRs of g/c-\ce{CH2CHCH2CN} and c/t-\ce{CH3CHCHCN}, although further refinement of chemical models is needed \citep{mallo_isomer-specific_2025}. In contrast, the HCOOH OSR requires a more complex network involving both gas-phase and grain-surface chemistry \citep{molpeceres_formic_2025}.While a comprehensive investigation of the conformer-specific chemistry of \ce{CH2CHCHO} is left for future work, existing gas–grain chemical networks can help to identify relevant formation and destruction pathways. We modeled the abundance of \ce{CH2CHCHO}, irrespective of isomeric form, using the \texttt{kida.2024} network \citep{wakelam_2024_2024} coupled with the \textsc{Nautilus} three-phase model \citep{ruaud_gas_2016}. Default parameters (\rev{whose basic details are grouped in Appendix \ref{app:a}}) were adopted except for the C/O ratio, which was varied to assess its effect. As shown in Figure~\ref{fig:model}, the model reasonably reproduces the observed abundance, with only a weak dependence on the C/O ratio, a frequent source of discrepancies between modeled and observed abundances in TMC-1. The predicted column density peaks at $\sim3.5\times10^{11}$ cm$^{-2}$ at times $\sim$6$\times$10$^{4}$ yr, earlier than the characteristic times obtained for TMC-1 \citep{agundez_tmc-1_2026}. Analysis of the formation rates shows that, at all relevant evolutionary times, nearly 100\% of \ce{CH2CHCHO} production proceeds through the neutral--neutral reaction
\begin{equation} \label{reac:1}
    \ce{C3H5 + O -> C2H3CHO + H},
\end{equation}
which is the oxygen analogue of the reaction investigated for the formation of propenethial by \citet{cabezas_discovery_2025} and \citet{molpeceres_laboratory_2026}. In the latter study, we found that this reaction pathway selectively produces the \textit{s-trans} conformer. If reaction~\ref{reac:1} shows a similar stereochemical preference, additional formation pathways must be invoked to account not only for the high observed OSR, but also for the presence of sc-\ce{CH2CHCHO} itself.

Equally, if not more important than formation pathways are the processes governing the destruction and interconversion of stereoisomers, which may strongly influence the OSRs. The relative dipole moment principle (RDP; \citealt{Shingledecker2020,shingledecker_grain-surface_2022,remijan_missing_2025}) predicts preferential destruction of more polar stereoisomers through enhanced ion--neutral capture rates, while the sequential acid--base (SAB) mechanism (\citealt{laas_contributions_2011,garcia_de_la_concepcion_sequential_2023}) allows isomeric redistribution through protonation followed by neutralization. Additionally, quantum tunneling (\citealt{GarciadelaConcepcion2021,garcia_de_la_concepcion_sequential_2023,molpeceres_laboratory_2026,del_valle_atom_2026}) enables interconversion between stereoisomers even at the low temperatures of molecular clouds, with rates strongly dependent on the potential energy barriers and relative isomer stabilities. Grain-surface chemistry further introduces formation, destruction, and isomerization pathways that may be particularly relevant following desorption \citep{molpeceres_formic_2025}. Thus, the OSR observed for \ce{CH2CHCHO} likely reflects the interplay of these mechanisms rather than a single dominant process. A comprehensive assessment of their relative contributions will be addressed in a future study, building upon the detection reported here.

\section{Conclusions}

The detection of $s$-cis-\ce{CH2CHCHO} in TMC-1 extends the
inventory of interstellar conformers and reveals a substantial
population of the higher-energy form. Its abundance relative
to $s$-trans acrolein is far from equilibrium predictions.
Although exploratory chemical models reproduce the total acrolein
abundance reasonably well, conformer-specific chemistry is
needed to explain the OSR of 0.45. Future studies should
delve into the stereochemical outcome of gas and grain phase formation, and subsequent destruction and interconversion processes.

\begin{acknowledgements}

Based on observations carried out with the Yebes 40m telescope
(projects 19A003, 20A014, 20D023, 21A011, 21D005, and 23A024).
The 40m radio telescope at Yebes Observatory is operated by the Spanish Geographic Institute (IGN; Ministerio de Transportes y Movilidad Sostenible). This work is supported by ERC grant No.
101218790 (Isocosmos) funded by the European Union. Views and opinions expressed are however those of the author(s) only
and do not necessarily reflect those of the European Union or
the European Research Council Executive Agency. Neither the
European Union nor the granting authority can be held responsible for them. We acknowledge support from the grants PID2024-156686NB-I00, PID2025-171315NB-I00 and RYC2022-035442-I funded by MICIU/AEI/10.13039/501100011033 and ESF+. G.M also acknowledges the support from project 20245AT016 (Proyectos
Intramurales CSIC). N.M., G.E. and B.T. acknowledge support from the Spanish grant PID2022-137980NB-I00, funded by MCIN/AEI/10.13039/501100011033/FEDER UE and PID2023-147545NB-I00.

\end{acknowledgements}

\bibliographystyle{aa}
\bibliography{bib_nuria}

@inproceedings{cernicharo_quijote_2022,
	title = {The {QUIJOTE} line survey of {TMC}-1},
	volume = {265},
	copyright = {http://creativecommons.org/licenses/by/4.0/},
	issn = {2100-014X},
	url = {https://www.epj-conferences.org/10.1051/epjconf/202226500041},
	doi = {10.1051/epjconf/202226500041},
	urldate = {2024-09-25},
	booktitle = {EPJ Web Conf.},
	author = {Cernicharo, José and Agúndez, Marcelino and Cabezas, Carlos and Marcelino, Nuria and Tercero, Belén and Pardo, Juan Ramón and Fuentetaja, Raúl and De Vicente, Pablo},
	editor = {Bouscasse, L. and Kramer, C. and Gueth, F.},
	year = {2022},
	pages = {00041},
}

@article{wakelam_efficiency_2021,
	title = {Efficiency of non-thermal desorptions in cold-core conditions: {Testing} the sputtering of grain mantles induced by cosmic rays},
	volume = {652},
	copyright = {https://creativecommons.org/licenses/by/4.0},
	issn = {0004-6361, 1432-0746},
	shorttitle = {Efficiency of non-thermal desorptions in cold-core conditions},
	url = {https://www.aanda.org/10.1051/0004-6361/202039855},
	doi = {10.1051/0004-6361/202039855},
	urldate = {2025-09-10},
	journal = {Astronomy \& Astrophysics},
	author = {Wakelam, V. and Dartois, E. and Chabot, M. and Spezzano, S. and Navarro-Almaida, D. and Loison, J.-C. and Fuente, A.},
	month = aug,
	year = {2021},
	pages = {A63},
}

@article{tercero_yebes_2021,
	title = {Yebes 40 m radio telescope and the broad band {Nanocosmos} receivers at 7 mm and 3 mm for line surveys},
	volume = {645},
	copyright = {https://www.edpsciences.org/en/authors/copyright-and-licensing},
	issn = {0004-6361, 1432-0746},
	url = {https://www.aanda.org/10.1051/0004-6361/202038701},
	doi = {10.1051/0004-6361/202038701},
	urldate = {2026-10-08},
	journal = {A\&A},
	author = {Tercero, F. and López-Pérez, J. A. and Gallego, J. D. and Beltrán, F. and García, O. and Patino-Esteban, M. and López-Fernández, I. and Gómez-Molina, G. and Diez, M. and García-Carreño, P. and Malo, I. and Amils, R. and Serna, J. M. and Albo, C. and Hernández, J. M. and Vaquero, B. and González-García, J. and Barbas, L. and López-Fernández, J. A. and Bujarrabal, V. and Gómez-Garrido, M. and Pardo, J. R. and Santander-García, M. and Tercero, B. and Cernicharo, J. and De Vicente, P.},
	month = jan,
	year = {2021},
	pages = {A37},
}

@article{agundez_tmc-1_2026,
	title = {{TMC}-1: {Probing} the {Onset} of {Chemical} {Complexity} in {Space}},
	volume = {10},
	copyright = {https://creativecommons.org/licenses/by/4.0/},
	issn = {2472-3452, 2472-3452},
	shorttitle = {{TMC}-1},
	url = {https://pubs.acs.org/doi/10.1021/acsearthspacechem.6c00026},
	doi = {10.1021/acsearthspacechem.6c00026},
	language = {en},
	number = {5},
	urldate = {2026-10-08},
	journal = {ACS Earth Space Chem.},
	author = {Agúndez, Marcelino and Cernicharo, José},
	month = may,
	year = {2026},
	pages = {1345--1359},
}

@article{cernicharo_discovery_2022,
	title = {Discovery of five cyano derivatives of propene with the {QUIJOTE} line survey},
	volume = {663},
	copyright = {https://creativecommons.org/licenses/by/4.0},
	issn = {0004-6361, 1432-0746},
	url = {https://www.aanda.org/10.1051/0004-6361/202244255},
	doi = {10.1051/0004-6361/202244255},
	urldate = {2026-03-01},
	journal = {A\&A},
	author = {Cernicharo, J. and Fuentetaja, R. and Cabezas, C. and Agúndez, M. and Marcelino, N. and Tercero, B. and Pardo, J. R. and De Vicente, P.},
	month = jul,
	year = {2022},
	pages = {L5},
}

@article{mcguire_early_2020,
	title = {Early {Science} from {GOTHAM}: {Project} {Overview}, {Methods}, and the {Detection} of {Interstellar} {Propargyl} {Cyanide} ({HCCCH}$_{\textrm{2}}$ {CN}) in {TMC}-1},
	volume = {900},
	issn = {2041-8205, 2041-8213},
	shorttitle = {Early {Science} from {GOTHAM}},
	url = {https://iopscience.iop.org/article/10.3847/2041-8213/aba632},
	doi = {10.3847/2041-8213/aba632},
	number = {1},
	urldate = {2026-10-21},
	journal = {ApJL},
	author = {McGuire, Brett A. and Burkhardt, Andrew M. and Loomis, Ryan A. and Shingledecker, Christopher N. and Lee, Kin Long Kelvin and Charnley, Steven B. and Cordiner, Martin A. and Herbst, Eric and Kalenskii, Sergei and Momjian, Emmanuel and Willis, Eric R. and Xue, Ci and Remijan, Anthony J. and McCarthy, Michael C.},
	month = sep,
	year = {2020},
	pages = {L10},
}

@article{cernicharo_discovery_2026,
	title = {Discovery of two new isomers of cyanoacenaphthylene ({C}$_{\textrm{12}}$ {H}$_{\textrm{7}}$ {CN}) in the {Taurus} molecular cloud 1 with the {QUIJOTE} line survey},
	volume = {705},
	copyright = {https://creativecommons.org/licenses/by/4.0},
	issn = {0004-6361, 1432-0746},
	url = {https://www.aanda.org/10.1051/0004-6361/202557893},
	doi = {10.1051/0004-6361/202557893},
	urldate = {2026-10-12},
	journal = {A\&A},
	author = {Cernicharo, J. and Tercero, B. and Marcelino, N. and López-Pérez, J. A. and Gallego, J. D. and Tercero, F. and Esplugues, G. and Cabezas, C. and Agúndez, M. and Limeres, C. and Steber, A. L. and Pérez, D. and Pérez, C. and Lesarri, A. and De Vicente, P.},
	month = jan,
	year = {2026},
	pages = {L7},
}

@article{Cernicharo2021,
	title = {Discovery of benzyne, o-{C6H4}, in {TMC}-1 with the {QUIJOTE} line survey⋆},
	volume = {652},
	url = {https://doi.org/10.1051/0004-6361/202141660},
	journal = {A\&A},
	author = {Cernicharo, J and Agúndez, M and Kaiser, R I and Cabezas, C and Tercero, B and Marcelino, N and Pardo, J R and de Vicente, P},
	year = {2021},
	pages = {L9},
}

@article{mccarthy_exhaustive_2020,
	title = {Exhaustive {Product} {Analysis} of {Three} {Benzene} {Discharges} by {Microwave} {Spectroscopy}},
	volume = {124},
	copyright = {https://doi.org/10.15223/policy-029},
	issn = {1089-5639, 1520-5215},
	url = {https://pubs.acs.org/doi/10.1021/acs.jpca.0c02919},
	doi = {10.1021/acs.jpca.0c02919},
	language = {en},
	number = {25},
	urldate = {2026-10-11},
	journal = {J. Phys. Chem. A},
	author = {McCarthy, Michael C. and Lee, Kin Long Kelvin and Carroll, P. Brandon and Porterfield, Jessica P. and Changala, P. Bryan and Thorpe, James H. and Stanton, John F.},
	month = jun,
	year = {2020},
	pages = {5170--5181},
}

@article{daly_comprehensive_2015,
	title = {{COMPREHENSIVE} {ANALYSIS} {OF} {PREBIOTIC} {PROPENAL} {UP} {TO} 660 {GHz}},
	volume = {218},
	copyright = {http://iopscience.iop.org/info/page/text-and-data-mining},
	issn = {1538-4365},
	url = {https://iopscience.iop.org/article/10.1088/0067-0049/218/2/30},
	doi = {10.1088/0067-0049/218/2/30},
	number = {2},
	urldate = {2024-09-02},
	journal = {ApJS},
	author = {Daly, A. M. and Bermúdez, C. and Kolesniková, L. and Alonso, J. L.},
	month = jun,
	year = {2015},
	pages = {30},
}

@article{sanz-novo_conformational_2025,
	title = {Conformational isomerism of methyl formate: {New} detections of the higher-energy \textit{trans} conformer and theoretical insights},
	volume = {698},
	copyright = {https://creativecommons.org/licenses/by/4.0},
	issn = {0004-6361, 1432-0746},
	shorttitle = {Conformational isomerism of methyl formate},
	url = {https://www.aanda.org/10.1051/0004-6361/202554686},
	doi = {10.1051/0004-6361/202554686},
	urldate = {2026-02-16},
	journal = {A\&A},
	author = {Sanz-Novo, M. and Molpeceres, G. and Rivilla, V. M. and Jimenez-Serra, I.},
	month = jun,
	year = {2025},
	pages = {A36},
}

@article{molpeceres_hydrogen_2022,
	title = {Hydrogen abstraction reactions in formic and thioformic acid isomers by hydrogen and deuterium atoms},
	volume = {663},
	issn = {0004-6361, 1432-0746},
	url = {https://www.aanda.org/10.1051/0004-6361/202243366},
	doi = {10.1051/0004-6361/202243366},
	urldate = {2022-12-13},
	journal = {A\&A},
	author = {Molpeceres, G. and Jiménez-Serra, I. and Oba, Y. and Nguyen, T. and Watanabe, N. and de la Concepción, J. García and Maté, B. and Oliveira, R. and Kästner, J.},
	month = jul,
	year = {2022},
	pages = {A41},
}

@article{goicoechea_ionization_2009,
	title = {The ionization fraction gradient across the {Horsehead} edge: an archetype for molecular clouds},
	volume = {498},
	issn = {0004-6361, 1432-0746},
	shorttitle = {The ionization fraction gradient across the {Horsehead} edge},
	url = {http://www.aanda.org/10.1051/0004-6361/200811496},
	doi = {10.1051/0004-6361/200811496},
	number = {3},
	urldate = {2026-10-11},
	journal = {A\&A},
	author = {Goicoechea, J. R. and Pety, J. and Gerin, M. and Hily-Blant, P. and Le Bourlot, J.},
	month = may,
	year = {2009},
	pages = {771--783},
}

@article{cuadrado_trans-cis_2016,
	title = {\textit{{Trans}-cis} molecular photoswitching in interstellar space},
	volume = {596},
	issn = {0004-6361, 1432-0746},
	url = {http://www.aanda.org/10.1051/0004-6361/201629913},
	doi = {10.1051/0004-6361/201629913},
	urldate = {2024-03-27},
	journal = {A\&A},
	author = {Cuadrado, S. and Goicoechea, J. R. and Roncero, O. and Aguado, A. and Tercero, B. and Cernicharo, J.},
	month = dec,
	year = {2016},
	pages = {L1},
}

@article{loison_interstellar_2017,
	title = {The interstellar chemistry of {C3H} and {C3H2} isomers},
	volume = {470},
	issn = {0035-8711, 1365-2966},
	url = {https://academic.oup.com/mnras/article/470/4/4075/3983227},
	doi = {10.1093/mnras/stx1265},
	language = {en},
	number = {4},
	urldate = {2026-01-01},
	journal = {MNRAS},
	author = {Loison, Jean-Christophe and Agúndez, Marcelino and Wakelam, Valentine and Roueff, Evelyne and Gratier, Pierre and Marcelino, Núria and Reyes, Dianailys Nuñez and Cernicharo, José and Gerin, Maryvonne},
	month = oct,
	year = {2017},
	pages = {4075--4088},
}

@article{hacar_hcn--hnc_2020,
	title = {{HCN}-to-{HNC} intensity ratio: a new chemical thermometer for the molecular {ISM}},
	volume = {635},
	copyright = {https://www.edpsciences.org/en/authors/copyright-and-licensing},
	issn = {0004-6361, 1432-0746},
	shorttitle = {{HCN}-to-{HNC} intensity ratio},
	url = {https://www.aanda.org/10.1051/0004-6361/201936516},
	doi = {10.1051/0004-6361/201936516},
	urldate = {2024-09-23},
	journal = {A\&A},
	author = {Hacar, A. and Bosman, A. D. and Van Dishoeck, E. F.},
	month = mar,
	year = {2020},
	pages = {A4},
}

@article{del_valle_atom_2026,
	title = {Atom {Addition} {Formation} of {Thionylimide} ({HNSO}) on {Interstellar} {Dust} {Grains}: {Chemical} {Routes} {Requiring} {Oxygen} and {Nitrogen} {Atom} {Surface} {Diffusion}},
	copyright = {https://creativecommons.org/licenses/by/4.0/},
	issn = {2472-3452, 2472-3452},
	shorttitle = {Atom {Addition} {Formation} of {Thionylimide} ({HNSO}) on {Interstellar} {Dust} {Grains}},
	url = {https://pubs.acs.org/doi/10.1021/acsearthspacechem.5c00360},
	doi = {10.1021/acsearthspacechem.5c00360},
	language = {en},
	urldate = {2026-02-20},
	journal = {ACS Earth Space Chem.},
	author = {Del Valle, Juan Carlos and Sanz-Novo, Miguel and Kästner, Johannes and Furuya, Kenji and Rivilla, Víctor M. and Martín-Doménech, Rafael and Molpeceres, Germán},
	month = feb,
	year = {2026},
  volume = {10},
  number = {3},
  pages = {742--756},
}

@article{GarciadelaConcepcion2021,
	title = {The {Origin} of the {E}/{Z} {Isomer} {Ratio} of {Imines} in the {Interstellar} {Medium}},
	volume = {912},
	issn = {2041-8205},
	url = {http://dx.doi.org/10.3847/2041-8213/abf650},
	doi = {10.3847/2041-8213/abf650},
	number = {1},
	journal = {ApJL},
	publisher = {American Astronomical Society},
	author = {García de la Concepción, Juan and Jiménez-Serra, Izaskun and Carlos Corchado, José and Rivilla, Víctor M. and Martín-Pintado, Jesús},
	year = {2021},
	pages = {L6},
}

@article{laas_contributions_2011,
	title = {{CONTRIBUTIONS} {FROM} {GRAIN} {SURFACE} {AND} {GAS} {PHASE} {CHEMISTRY} {TO} {THE} {FORMATION} {OF} {METHYL} {FORMATE} {AND} {ITS} {STRUCTURAL} {ISOMERS}},
	volume = {728},
	issn = {0004-637X, 1538-4357},
	url = {https://iopscience.iop.org/article/10.1088/0004-637X/728/1/71},
	doi = {10.1088/0004-637X/728/1/71},
	number = {1},
	urldate = {2024-09-03},
	journal = {ApJ},
	author = {Laas, Jacob C. and Garrod, Robin T. and Herbst, Eric and Widicus Weaver, Susanna L.},
	month = feb,
	year = {2011},
	pages = {71},
}

@article{garcia_de_la_concepcion_sequential_2023,
	title = {A sequential acid-base mechanism in the interstellar medium: {The} emergence of cis-formic acid in dark molecular clouds},
	volume = {675},
	copyright = {https://creativecommons.org/licenses/by/4.0},
	issn = {0004-6361, 1432-0746},
	shorttitle = {A sequential acid-base mechanism in the interstellar medium},
	url = {https://www.aanda.org/10.1051/0004-6361/202243966},
	doi = {10.1051/0004-6361/202243966},
	urldate = {2024-03-27},
	journal = {A\&A},
	author = {García De La Concepción, J. and Jiménez-Serra, I. and Corchado, J. C. and Molpeceres, G. and Martínez-Henares, A. and Rivilla, V. M. and Colzi, L. and Martín-Pintado, J.},
	month = jul,
	year = {2023},
	pages = {A109},
}

@article{ruaud_gas_2016,
	title = {Gas and grain chemical composition in cold cores as predicted by the {Nautilus} three-phase model},
	volume = {459},
	issn = {13652966},
	url = {https://academic.oup.com/mnras/article-lookup/doi/10.1093/mnras/stw887},
	doi = {10.1093/mnras/stw887},
	number = {4},
	urldate = {2017-05-04},
	journal = {MNRAS},
	author = {Ruaud, Maxime and Wakelam, Valentine and Hersant, Franck},
	month = jul,
	year = {2016},
	pages = {3756--3767},
}

@article{remijan_missing_2025,
	title = {The {Missing} {Link} of {Sulfur} {Chemistry} in {TMC}-1: {The} {Detection} of \textit{c} -{C}$_{\textrm{3}}$ {H}$_{\textrm{2}}$ {S} from the {GOTHAM} {Survey}},
	volume = {982},
	issn = {0004-637X, 1538-4357},
	shorttitle = {The {Missing} {Link} of {Sulfur} {Chemistry} in {TMC}-1},
	url = {https://iopscience.iop.org/article/10.3847/1538-4357/adb84c},
	doi = {10.3847/1538-4357/adb84c},
	number = {2},
	urldate = {2026-10-11},
	journal = {ApJ},
	author = {Remijan, Anthony J. and Changala, P. Bryan and Xue, Ci and Yuan, Elsa Q. H. and Duffy, Miya and Scolati, Haley N. and Shingledecker, Christopher N. and Speak, Thomas H. and Cooke, Ilsa R. and Loomis, Ryan and Burkhardt, Andrew M. and Fried, Zachary T. P. and Wenzel, Gabi and Lipnicky, Andrew and McCarthy, Michael C. and McGuire, Brett A.},
	month = apr,
	year = {2025},
	pages = {191},
}

@article{shingledecker_grain-surface_2022,
	title = {Grain-{Surface} {Hydrogen}-{Addition} {Reactions} as a {Chemical} {Link} {Between} {Cold} {Cores} and {Hot} {Corinos}: {The} {Case} of {H}$_{\textrm{2}}$ {CCS} and {CH}$_{\textrm{3}}$ {CH}$_{\textrm{2}}$ {SH}},
	volume = {126},
	copyright = {https://doi.org/10.15223/policy-029},
	issn = {1089-5639, 1520-5215},
	shorttitle = {Grain-{Surface} {Hydrogen}-{Addition} {Reactions} as a {Chemical} {Link} {Between} {Cold} {Cores} and {Hot} {Corinos}},
	url = {https://pubs.acs.org/doi/10.1021/acs.jpca.2c01447},
	doi = {10.1021/acs.jpca.2c01447},
	language = {en},
	number = {32},
	urldate = {2026-10-11},
	journal = {J. Phys. Chem. A},
	author = {Shingledecker, Christopher N. and Banu, Tahamida and Kang, Yi and Wei, Hongji and Wandishin, Joseph and Nobis, Garrett and Jarvis, Virginia and Quinn, Faith and Quinn, Grace and Molpeceres, Germán and McCarthy, Michael C. and McGuire, Brett A. and Kästner, Johannes},
	month = aug,
	year = {2022},
	pages = {5343--5353},
}

@article{Shingledecker2020,
	title = {Isomers in {Interstellar} {Environments}. {I}. {The} {Case} of {Z}- and {E}-cyanomethanimine},
	volume = {897},
	issn = {1538-4357},
	doi = {10.3847/1538-4357/ab94b5},
	number = {2},
	journal = {ApJ},
	author = {Shingledecker, Christopher and Molpeceres, Germán and Rivilla, Víctor and Majumdar, Liton and Kästner, Johannes},
	year = {2020},
	pages = {158},
}

@article{wakelam_2024_2024,
	title = {The 2024 {KIDA} network for interstellar chemistry},
	volume = {689},
	copyright = {https://creativecommons.org/licenses/by/4.0},
	issn = {0004-6361, 1432-0746},
	url = {https://www.aanda.org/10.1051/0004-6361/202450606},
	doi = {10.1051/0004-6361/202450606},
	urldate = {2026-10-10},
	journal = {A\&A},
	author = {Wakelam, V. and Gratier, P. and Loison, J.-C. and Hickson, K. M. and Penguen, J. and Mechineau, A.},
	month = sep,
	year = {2024},
	pages = {A63},
}

@article{Lee2021,
	title = {Discovery of {Interstellar} trans -cyanovinylacetylene ({HC} ≡ {CCH} = {CHC} ≡ {N}) and vinylcyanoacetylene ({H} 2 {C} = {CHC} 3 {N}) in {GOTHAM} {Observations} of {TMC}-1},
	volume = {908},
	issn = {20418213},
	url = {http://dx.doi.org/10.3847/2041-8213/abdbb9},
	doi = {10.3847/2041-8213/abdbb9},
	number = {1},
	journal = {ApJ},
	publisher = {American Astronomical Society},
	author = {Lee, Kin Long Kelvin and Loomis, Ryan A. and Burkhardt, Andrew M. and Cooke, Ilsa R. and Xue, Ci and Siebert, Mark A. and Shingledecker, Christopher N. and Remijan, Anthony and Charnley, Steven B. and McCarthy, Michael C. and McGuire, Brett A.},
	year = {2021},
	pages = {L11},
}

@article{mallo_isomer-specific_2025,
	title = {Isomer-{Specific} {Branching} {Ratios} in the {Formation} of {Cyanopropene} ({C}$_{\textrm{3}}$ {H}$_{\textrm{5}}$ {CN}) through the {C}$_{\textrm{3}}$ {H}$_{\textrm{6}}$ + {CN} {Reaction} under {Interstellar} {Conditions}},
	volume = {9},
	copyright = {https://creativecommons.org/licenses/by/4.0/},
	issn = {2472-3452, 2472-3452},
	url = {https://pubs.acs.org/doi/10.1021/acsearthspacechem.5c00274},
	doi = {10.1021/acsearthspacechem.5c00274},
	language = {en},
	number = {11},
	urldate = {2025-11-28},
	journal = {ACS Earth Space Chem.},
	author = {Mallo, María and Agúndez, Marcelino and Cabezas, Carlos and Cernicharo, José and Molpeceres, Germán},
	month = nov,
	year = {2025},
	pages = {2747--2758},
}

@article{molpeceres_formic_2025,
	title = {Formic acid isomerism in dark clouds: {Detection} of cis-formic acid in {TMC}-1 with the {QUIJOTE} line survey and astrochemical modeling},
	volume = {703},
	copyright = {https://creativecommons.org/licenses/by/4.0},
	issn = {0004-6361, 1432-0746},
	shorttitle = {Formic acid isomerism in dark clouds},
	url = {https://www.aanda.org/10.1051/0004-6361/202555969},
	doi = {10.1051/0004-6361/202555969},
	urldate = {2026-01-04},
	journal = {A\&A},
	author = {Molpeceres, G. and Agúndez, M. and Mallo, M. and Cabezas, C. and Sanz-Novo, M. and Rivilla, V. M. and García De La Concepción, J. and Jiménez-Serra, I. and Cernicharo, J.},
	month = nov,
	year = {2025},
	pages = {A164},
}

@article{rivilla_interstellar_2026,
	title = {Interstellar {Stereoisomerism}},
	copyright = {https://creativecommons.org/licenses/by/4.0/},
	issn = {2472-3452, 2472-3452},
	url = {https://pubs.acs.org/doi/10.1021/acsearthspacechem.6c00001},
	doi = {10.1021/acsearthspacechem.6c00001},
	language = {en},
	urldate = {2026-05-11},
	journal = {ACS Earth Space Chem.},
	author = {Rivilla, Víctor M. and Sanz-Novo, Miguel and San Andrés, David},
	month = apr,
	year = {2026},
	pages = {1155--1173},
  volume = {10},
  number = {5},
}

@ARTICLE{1987A&A...176..299C,
       author = {{Cernicharo}, J. and {Guelin}, M.},
        title = "{The physical and chemical state of HCL2.}",
      journal = {A\&A},
         year = 1987,
        month = apr,
       volume = {176},
        pages = {299-316},
       adsurl = {https://ui.adsabs.harvard.edu/abs/1987A&A...176..299C}
}

@article{cernicharo_spatial_2023,
	title = {The spatial distribution of an aromatic molecule, {C}$_{\textrm{6}}$ {H}$_{\textrm{5}}$ {CN}, in the cold dark cloud {TMC}-1},
	volume = {674},
	copyright = {https://creativecommons.org/licenses/by/4.0},
	issn = {0004-6361, 1432-0746},
	url = {https://www.aanda.org/10.1051/0004-6361/202346722},
	doi = {10.1051/0004-6361/202346722},
	urldate = {2026-10-09},
	journal = {A\&A},
	author = {Cernicharo, J. and Tercero, B. and Marcelino, N. and Agúndez, M. and De Vicente, P.},
	month = jun,
	year = {2023},
	pages = {L4},
}

@article{fosse_molecular_2001,
	title = {Molecular {Carbon} {Chains} and {Rings} in {TMC}‐1},
	volume = {552},
	issn = {0004-637X, 1538-4357},
	url = {https://iopscience.iop.org/article/10.1086/320471},
	doi = {10.1086/320471},
	language = {en},
	number = {1},
	urldate = {2026-10-09},
	journal = {ApJ},
	author = {Fosse, David and Cernicharo, Jose and Gerin, Maryvonne and Cox, Pierre},
	month = may,
	year = {2001},
	pages = {168--174},
}

@inproceedings{cernicharo_laboratory_2012,
	title = {Laboratory astrophysics and astrochemistry in the {Herschel}/{ALMA} era},
	volume = {58},
	url = {http://www.eas-journal.org/10.1051/eas/1258040},
	doi = {10.1051/eas/1258040},
	urldate = {2026-07-06},
	booktitle = {EAS Publ. Ser.},
	series = {European Conference on Laboratory Astrophysics (ECLA 2011)},
	author = {Cernicharo, J.},
	editor = {Stehlé, C. and Joblin, C. and d’Hendecourt, L.},
	year = {2012},
	pages = {251--261},
}

@article{marcelino_first_2023,
	title = {First detection of the {HSO} radical in space},
	volume = {674},
	copyright = {https://creativecommons.org/licenses/by/4.0},
	issn = {0004-6361, 1432-0746},
	url = {https://www.aanda.org/10.1051/0004-6361/202346935},
	doi = {10.1051/0004-6361/202346935},
	urldate = {2026-10-09},
	journal = {A\&A},
	author = {Marcelino, N. and Puzzarini, C. and Agúndez, M. and Fuentetaja, R. and Tercero, B. and De Vicente, P. and Cernicharo, J.},
	month = jun,
	year = {2023},
	pages = {L13},
}

@article{molpeceres_laboratory_2026,
	title = {Laboratory spectroscopy, theoretical characterization, and astronomical search for \textit{syn} -propenethial ({CH2CHCHS})},
	volume = {551},
	copyright = {https://creativecommons.org/licenses/by/4.0/},
	issn = {0035-8711, 1365-2966},
	url = {https://academic.oup.com/mnras/article/doi/10.1093/mnras/stag1436/8747103},
	doi = {10.1093/mnras/stag1436},
	language = {en},
	number = {1},
	urldate = {2026-09-16},
	journal = {MNRAS},
	author = {Molpeceres, Germán and Cabezas, Carlos and Agúndez, Marcelino and Mallo, María and Endo, Yasuki and Kolesniková, Lucie and Esplugues, Gisela and Cernicharo, José},
	month = aug,
	year = {2026},
	pages = {stag1436},
}

@article{esplugues_first_2025,
	title = {First detection of {HS}$_{\textrm{2}}$ in a cold dark cloud},
	volume = {699},
	copyright = {https://creativecommons.org/licenses/by/4.0},
	issn = {0004-6361, 1432-0746},
	url = {https://www.aanda.org/10.1051/0004-6361/202555384},
	doi = {10.1051/0004-6361/202555384},
	urldate = {2026-10-09},
	journal = {A\&A},
	author = {Esplugues, G. and Agúndez, M. and Molpeceres, G. and Tercero, B. and Cabezas, C. and Marcelino, N. and Fuentetaja, R. and Cernicharo, J.},
	month = jul,
	year = {2025},
	pages = {L2},
}

@ARTICLE{2020A&A...642L...8C,
       author = {{Cernicharo}, J. and {Marcelino}, N. and {Ag{\'u}ndez}, M. and {Berm{\'u}dez}, C. and {Cabezas}, C. and {Tercero}, B. and {Pardo}, J.~R.},
        title = "{Discovery of HC$_{4}$NC in TMC-1: A study of the isomers of HC$_{3}$N, HC$_{5}$N, and HC$_{7}$N}",
      journal = {A\&A},
         year = 2020,
        month = oct,
       volume = {642},
          eid = {L8},
        pages = {L8},
          doi = {10.1051/0004-6361/202039274},
       adsurl = {https://ui.adsabs.harvard.edu/abs/2020A&A...642L...8C}
}

@article{cernicharo_discovery_2021,
	title = {Discovery of {HCCCO} and {C}$_{\textrm{5}}$ {O} in {TMC}-1 with the {QUIJOTE} line survey},
	volume = {656},
	copyright = {https://www.edpsciences.org/en/authors/copyright-and-licensing},
	issn = {0004-6361, 1432-0746},
	url = {https://www.aanda.org/10.1051/0004-6361/202142634},
	doi = {10.1051/0004-6361/202142634},
	urldate = {2026-10-09},
	journal = {A\&A},
	author = {Cernicharo, J. and Agúndez, M. and Cabezas, C. and Tercero, B. and Marcelino, N. and Fuentetaja, R. and Pardo, J. R. and De Vicente, P.},
	month = dec,
	year = {2021},
	pages = {L21},
}

\begin{appendix}

\twocolumn[{%
\section{Parameters of the \textsc{Nautilus} chemical model} \label{app:a}

\begin{minipage}{\textwidth}
\centering
\captionof{table}{Basic physical conditions and initial abundances adopted in the chemical models.}
\label{tab:nautilus-conditions}
\small

\begin{minipage}[t]{0.61\textwidth}
\centering
\begin{tabular}{lr}
\hline\hline
\multicolumn{2}{c}{Physical conditions} \\
\hline
Network & \texttt{kida.2024} \\
$n_{\rm H}$ & $2.0\times10^{4}$ cm$^{-3}$ \\
$T_{\rm gas}$ & 10 K \\
$T_{\rm dust}$ & 10 K \\
$A_V$ & 15 mag \\
$\zeta$ & $3.77\times10^{-17}$ s$^{-1}$ \\
C/O & 0.71, 1.00, 1.25 \\
\hline
\end{tabular}
\end{minipage}\hfill
\begin{minipage}[t]{0.36\textwidth}
\centering
\begin{tabular}{lr}
\hline\hline
Species & $n(i)/n_{\rm H}$ \\
\hline
\ce{H2} & 0.5 \\
He & $9.0\times10^{-2}$ \\
\ce{C+} & $1.7\times10^{-4}$ \\
N & $6.2\times10^{-5}$ \\
O & $(2.40, 1.70, 1.36)\times10^{-4}$ \\
\ce{S+} & $8.0\times10^{-8}$ \\
\ce{Si+} & $8.0\times10^{-9}$ \\
\ce{Fe+} & $3.0\times10^{-9}$ \\
\ce{Na+} & $2.0\times10^{-9}$ \\
\ce{Mg+} & $7.0\times10^{-9}$ \\
\ce{P+} & $2.0\times10^{-10}$ \\
\ce{Cl+} & $1.0\times10^{-9}$ \\
F & $6.68\times10^{-9}$ \\
\hline
\end{tabular}
\end{minipage}

\tablefoot{
Abundances are relative to $n_{\rm H}$. The cosmic ray ionization rate ($\zeta$) is calculated with respect to  $A_V$ according to the prescription in \citet{wakelam_efficiency_2021}. Further parameters can be consulted in the \texttt{parameter} file accompanying the release of \textsc{Nautilus}. 
}
\end{minipage}}]

\end{appendix}

\end{document}